\documentclass[acmsmall,screen]{acmart}
\usepackage{soul}
\usepackage{hyperref}
\usepackage{wrapfig}
\usepackage{multirow}
\usepackage{listings}
\usepackage{fancybox}
\usepackage{enumitem}
\usepackage{graphicx}
\usepackage{color}
\usepackage{xcolor}
\usepackage{fancyvrb}
\usepackage{array}
\usepackage{amsmath}
\usepackage{xspace}
\usepackage{url}
\usepackage{tikz}
\usepackage{caption}
\usepackage{pgfplots}
\usepackage{balance}
\usepackage{subcaption}
\pgfplotsset{compat=1.16}
\usepackage{pgf-pie}
\usepackage{float}
\usetikzlibrary{pgfplots.statistics,calc}
\usepackage{algorithm}
\usepackage{algorithmic}
\usepackage{adjustbox}
\usepackage{makecell}

\usepackage{tcolorbox}
\makeatletter
\newcommand{\mybox}[1]{%
	\setbox0=\hbox{#1}%
	\setlength{\@tempdima}{\dimexpr\wd0+13pt}%
	\begin{tcolorbox}[boxrule=0.5pt, colback=white, arc=4pt,
		left=6pt,right=6pt,top=6pt,bottom=6pt,boxsep=0pt]
		#1
	\end{tcolorbox}
}

\definecolor{songcolor}{RGB}{191,191,191}

\AtBeginDocument{%
  }

\setcopyright{acmcopyright}
\copyrightyear{2024}
\acmYear{2024}
\acmDOI{XXXXXXX.XXXXXXX}
\acmPrice{15.00}
\acmISBN{978-1-4503-XXXX-X/18/06}
\usepackage{listings}
\usepackage{xcolor}
\usepackage{balance}
\usepackage{tcolorbox}
\newcounter{o}
\begin{document}

\definecolor{codegreen}{rgb}{0,0.6,0}
\definecolor{codegray}{rgb}{0.5,0.5,0.5}
\definecolor{codepurple}{rgb}{0.58,0,0.82}
\definecolor{backcolour}{rgb}{0.95,0.95,0.92}

\lstdefinestyle{mystyle}{
    backgroundcolor=\color{backcolour},   
    commentstyle=\color{codegreen},
    keywordstyle=\color{magenta},
    numberstyle=\tiny\color{codegray},
    stringstyle=\color{codepurple},
    basicstyle=\ttfamily\footnotesize,
    breakatwhitespace=false,         
    breaklines=true,                 
    captionpos=b,                    
    keepspaces=true,                 
    numbers=left,                    
    numbersep=5pt,                  
    showspaces=false,                
    showstringspaces=false,
    showtabs=false,                  
    tabsize=2
}

\lstset{style=mystyle}

\title[]{Demystifying Errors in LLM Reasoning Traces: An Empirical Study of Code Execution Simulation}

\author{Mohammad Abdollahi}
\affiliation{%
  \institution{York University}
  \streetaddress{4700 Keele St.}
  \city{North York}
  \country{Canada}}
\email{moham98@yorku.ca}

\author{Khandaker Rifah Tasnia}
\affiliation{%
  \institution{Concordia University}
  \streetaddress{1455 de Maisonneuve Blvd}
  \city{Montreal}
  \country{Canada}}
\email{khandakerrifah.tasnia@mail.concordia.ca}

\author{Soumit Kanti Saha}
\affiliation{%
  \institution{Concordia University}
  \streetaddress{1455 de Maisonneuve Blvd}
  \city{Montreal}
  \country{Canada}}
\email{saha.soumit884@gmail.com}

\author{Jinqiu Yang}
\affiliation{%
  \institution{Concordia University}
  \streetaddress{1455 de Maisonneuve Blvd}
  \city{Montreal}
  \country{Canada}}
\email{jinqiu.yang@concordia.ca}

\author{Song Wang}
\affiliation{%
  \institution{York University}
  \streetaddress{4700 Keele St.}
  \city{North York}
  \country{Canada}}
\email{wangsong@yorku.ca}

\author{Hadi Hemmati}
\affiliation{%
  \institution{York University}
  \streetaddress{4700 Keele St.}
  \city{North York}
  \country{Canada}}
\email{hemmati@yorku.ca}

\begin{abstract}

Understanding a program’s runtime reasoning behavior, i.e., how intermediate states and control flows lead to final execution results, is essential for reliable code generation, debugging, and automated reasoning. While large language models (LLMs) have demonstrated impressive capabilities in predicting program outputs, most prior studies have focused on output accuracy and performance comparisons, treating reasoning as a black box. As a result, models often struggle with complex control logic or subtle semantic nuances, and little is known about the structure, quality, or failure modes of their reasoning traces.

Recent advances in reasoning LLMs, which explicitly generate intermediate reasoning steps before producing final answers, raise the expectation that models could not only produce correct code but also explain their reasoning in ways consistent with program semantics. This ability is particularly critical for tasks involving runtime behavior, where reasoning must capture both control flow and state changes. However, the lack of systematic evaluation of reasoning traces leaves open fundamental questions about how reasoning errors arise and how they influence execution correctness. 

To address this gap, we conduct the first empirical study on reasoning runtime behavior inference
with reasoning LLMs to systematically uncover and characterize the errors in their reasoning traces. 
For our research, we curate a benchmark constructed from two widely used datasets (i.e., HumanEval+ and LiveCodeBench), containing 427 code snippets. For each code snippet, we experiment with three different types of inputs, i.e., regular, edge, and invalid inputs. A total of 12 input values are selected per snippet, each paired with its ground-truth execution outcome. 
We evaluate four state-of-the-art reasoning LLMs, i.e., DeepSeek-R1, OpenAI o4-mini, Gemini 2.5 Flash, and Claude 4 Sonnet, and our experiment results show that the examined reasoning LLMs can achieve high accuracies ranging from 85\% to 98\% across all three input types. 
%We also observe that their accuracy drops sharply on complex code snippets, such as those with nested logic, recursion, or implicit state changes, compared to simpler ones. This highlights the persistent gap between surface-level token prediction and genuine semantic reasoning. 
Additionally, we analyzed the errors in reasoning traces produced by LLMs and developed a comprehensive taxonomy encompassing nine categories of inference errors in the code output inference tasks.  
% \song{mention you also analyze the failed inference and build a failure reason taxonomy}

We further investigate whether a tool-augmented reasoning approach can enhance the reasoning ability of LLMs. Using failures in the \textit{Computation Errors} category as a case study, our experiments show that this method successfully corrects 58\% of such errors, underscoring the promise of tool augmentation in improving LLM reasoning.

\end{abstract}




%\received{28 September 2024}
%\received[revised]{5 March 2025}
%\received[accepted]{16 April 2025}

\setcopyright{none} % to remove the copyright notice
\settopmatter{printacmref=false} % to remove the ACM Reference Format
\renewcommand\footnotetextcopyrightpermission[1]{}
%%
%% This command processes the author, affiliation, and title
\maketitle

\section{Introduction}
\label{sec:intro}

Large Language Models (LLMs) have become increasingly capable in software engineering tasks, including code generation, translation, summarization, and reasoning~\cite{abdollahi2025surveying,yang2025code}. 
Recent developments in reasoning LLMs, i.e., models that explicitly generate intermediate reasoning traces before producing a final answer, have raised expectations that these systems may not only produce correct code, but also explain their reasoning in a manner that aligns with program semantics~\cite{guo2025deepseek}. This capability is particularly relevant for tasks that require understanding a program’s runtime behavior, i.e., the sequence of operations and state changes that occur during execution~\cite{gu2024cruxeval}.

Prior work~\cite{liu2024codemind,chen2024reasoning,beger2025coconut,tang2025codereasoner,xie2025core,yang2025evaluating} on assessing LLMs’ capability to reason about the runtime behavior of program execution has predominantly examined performance differences across models. For example, CodeMind~\cite{liu2024codemind} evaluated a range of LLMs, including both general-purpose and code-specialized models, on code reasoning tasks such as semantic understanding and output prediction. REval~\cite{chen2024reasoning} assessed the reasoning ability and execution consistency of code LLMs, introducing an Incremental Consistency (IC) metric to measure how well a model maintains coherent reasoning across sequential tasks of increasing difficulty. CoCoNut~\cite{beger2025coconut} examined LLMs’ ability to trace program control flow when provided with relevant inputs, offering insights into their understanding of execution behavior. 

\begin{figure}[t!] 
\centering 
\includegraphics[width=\textwidth,height=0.5\textheight,keepaspectratio]{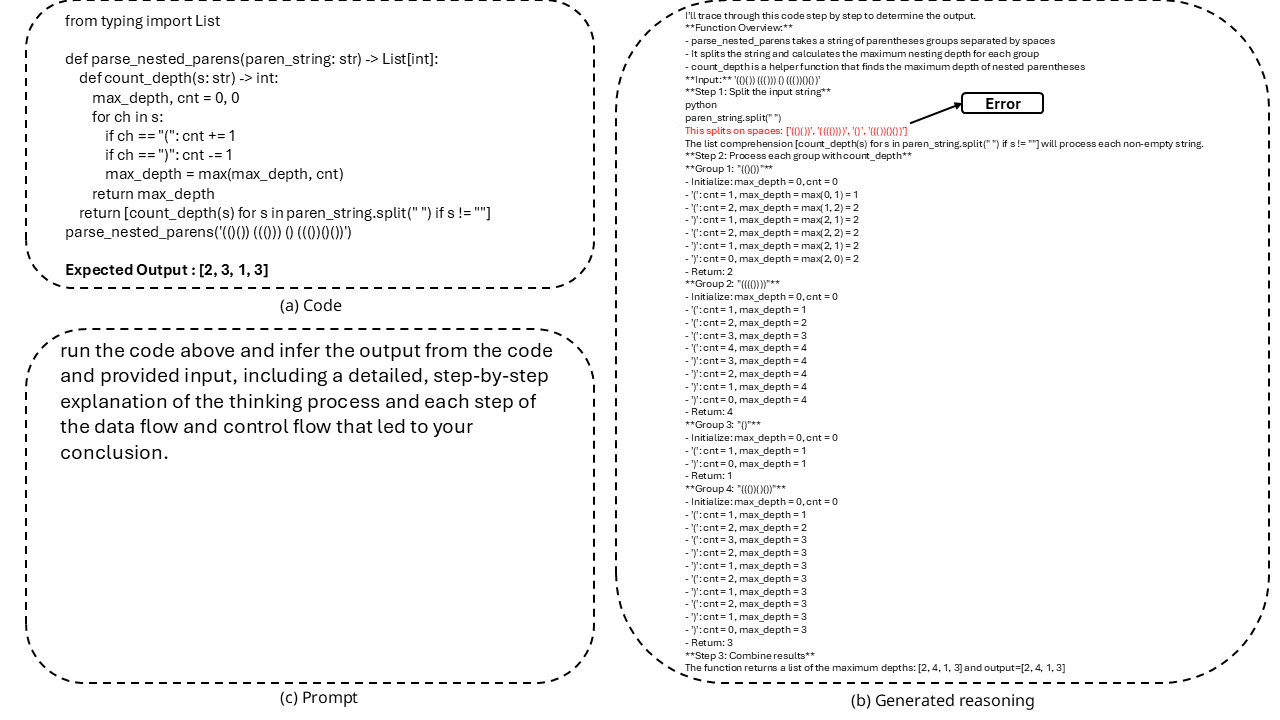} 
\caption{Outputs generated by the LLM: (a) code snippet, (b) corresponding reasoning, and (c) prompt.}
\label{fig:llm_reasoning_output_sample} 
\end{figure}

However, these studies largely focused on evaluating LLMs' output accuracy rather than on analyzing the reasoning traces generated by LLMs on code execution. As a result, little is known about the nature of LLM-generated reasoning traces, the types of errors they contain, and how these errors relate to the underlying program semantics, leaving a critical gap in understanding and improving LLMs’ true reasoning abilities. 
% \include{figs/fig1}
% \song{describe your Fig1}
Figure~\ref{fig:llm_reasoning_output_sample} illustrates a representative example of an LLM-generated reasoning trace for a code execution task. The figure presents the original code snippet, the corresponding prompt provided to the model, and the reasoning generated by the Claude 4 Sonnet. The LLM produces an incorrect output due to a misevaluation of Python’s \texttt{split()} function on strings, illustrating a reasoning error where the model fails to accurately capture the intended program semantics.

% \song{@Mohammad: can you show an example? You can use a simple example from HumanEval, as this is to illustrate what the reasoning generated by a reasoning LLM looks like, so a simple example should be fine, you can show the code and the reasoning generated by an LLM in two sub-figures of the same figure.}

To address this gap, we conduct the first empirical study on reasoning runtime behavior inference with reasoning LLMs to systematically uncover and characterize the errors in their reasoning traces. 
For conducting our analysis, we introduce a new benchmark constructed from two widely used program analysis datasets (i.e., HumanEval+ and LiveCodeBench), comprising 427 diverse code snippets. For every snippet, we experiment with three input categories, i.e., regular inputs that represent typical execution scenarios, edge inputs involving boundary conditions and rare execution paths, and invalid inputs, such as malformed data and type mismatches.  
%This setup enables us to evaluate not only reasoning LLMs’ baseline reasoning accuracy but also their robustness to corner cases and their ability to detect and handle execution errors. 
A total of 12 input values are selected per snippet, each paired with its ground-truth execution outcome. 

Using this new dataset, we evaluate four state-of-the-art reasoning LLMs, including DeepSeek-R1~\cite{deepseekai2025deepseekr1incentivizingreasoningcapability}, OpenAI o4-mini~\cite{openai2024openaio1card}, Gemini 2.5 Flash~\cite{comanici2025gemini}, and Claude 4 Sonnet~\cite{claude4} to explore their reasoning quality by answering the following research questions.
\begin{itemize}
   \item \textbf{RQ1  (Performance): How accurately can they infer execution outcomes across varying types of inputs?}

    \item \textbf{RQ2 (Failure Location): Which statements or code locations most frequently trigger errors in the LLMs’ reasoning process during code inference?}   

   \item \textbf{RQ3 (Failure Pattern): What are the common reasoning error patterns exhibited by reasoning LLMs during execution outcome inference?}

%   \item \textbf{RQ4 (Reasoning Difference): How do reasoning error patterns vary across different reasoning LLMs?}
   
\end{itemize}

Our empirical study shows that the evaluated reasoning LLMs (i.e., DeepSeek-R1, OpenAI o4-mini, Gemini 2.5 Flash, and Claude 4) achieve high accuracies ranging from 85\% to 98\% across all three input types. %(2) Reasoning LLMs' accuracy drops sharply on code with nested logic, recursion, or implicit state dependencies (e.g., variables updated across iterations). %These results highlight the persistent gap between surface-level token prediction and true semantic reasoning. 
We further analyze the reasoning errors produced by LLMs on code inference tasks, and we developed a structured taxonomy to categorize these errors. The taxonomy comprises nine high-level categories: \textit{Computation Errors}, \textit{Indexing Errors}, \textit{Control Flow Errors}, \textit{Skip Statements}, \textit{Misreporting Final Output}, \textit{Input Misread}, \textit{Misevaluation of Native API}, \textit{Hallucination}, and \textit{Lack of Verification/Logic Following}. This taxonomy provides a systematic framework for evaluating and understanding the limitations of LLMs in code inference.
% \song{describe your failure taxonomy results here}

To address reasoning errors in LLMs, we propose a tool-augmented reasoning approach that supplies reliable intermediate signals, reduces contextual noise, and enhances execution inference accuracy. We evaluate its effectiveness using failures in the \textit{Computation Errors} category as a case study. Our experiments show that this method successfully corrects 58\% of these errors, demonstrating the significant potential of tool augmentation to improve LLM reasoning performance. This paper makes the following contributions:

\begin{itemize}
\item We conduct the first large-scale comparative study of four state-of-the-art reasoning LLMs, i.e., GPT-o4 mini, Claude 4 Sonnet, DeepSeek R1, and Gemini 2.5 Flash, across diverse input types and complexity levels, uncovering critical weaknesses in how their intermediate reasoning traces fail to align with actual program execution outcomes.

\item  We perform a fine-grained analysis of reasoning traces, examining each failed case at two complementary levels: (1) Statement-level, which identifies the exact program statement where the model’s reasoning first diverges from the ground-truth execution, and (2) Trace-level, which characterizes broader breakdowns that affect the reasoning as a whole. 

\item We develop a taxonomy for understanding reasoning errors in LLMs, providing a structured lens to examine how models fail when performing code inference tasks. This taxonomy categorizes errors into nine high-level classes that capture both the nature of logical missteps and the program elements affected.
%We develop a taxonomy for understanding reasoning errors in LLMs, providing a structured framework to examine how models fail when performing code inference tasks. This taxonomy categorizes errors into nine high-level classes, some of which have subcategories, that capture both the nature of logical missteps and the program elements affected.

\item  We propose a tool-augmented approach that provides intermediate signals for reasoning LLMs, which can help correct 58\% reasoning errors in the \textit{Computation Errors} category.

%\item  Experiments show that our method consistently improves both reasoning trace faithfulness and execution outcome correctness across all evaluated LLMs.

\end{itemize}
%: How accurately can they predict execution outcomes across varying input complexities? What types of reasoning errors are most common? And to what extent do their intermediate reasoning traces reflect sound program semantics?

The remainder of this paper is organized as follows: Section~\ref{sec:bg} introduces the background of this work, Section~\ref{sec:approach} presents the process of our empirical study, 
Section~\ref{sec:results} shows the results, 
Section~\ref{sec:discussion} discusses the open questions that related to our approach, Section~\ref{sec:threats} discusses the threats to validity of this work, and Section~\ref{sec:conclusion} concludes this
work.

\section{Background and Related Work}
\label{sec:bg}

\subsection{Reasoning LLMs}
\label{sec:b0}

To understand how LLMs make decisions, researchers initially framed LLM reasoning as a prompting problem. Kojima et al.~\cite{kojima2022large} and Wei et al.~\cite{wei2022chain} introduced Chain-of-Thought (CoT) prompting, which instructs an LLM to produce a sequence of intermediate reasoning steps, significantly improving performance on complex reasoning benchmarks. Building on this idea, Self-Consistency~\cite{wang2022self} samples multiple CoT trajectories and aggregates their final answers to reduce stochastic errors. 
Subsequently, Tree-of-Thought (ToT)~\cite{yao2023tree} generalized the CoT paradigm by enabling exploration across coherent reasoning units (``thoughts'') as intermediate steps toward problem-solving. This approach searches over multiple partial solutions, trading computational cost for improved robustness. 
Beyond pure prompting, tool-augmented frameworks such as ReAct~\cite{yao2023react} and Toolformer~\cite{schick2023toolformer} interleave natural language reasoning with external actions, including information retrieval, calculator usage, or code execution. 

The above methods have been shown to generate accurate reasoning traces on many coding tasks~\cite{huang2023codecot,li2025structured,suzgun2022challenging,zhang2025they}. However, the reasoning produced in these approaches is often optimized for producing correct answers rather than being evaluated or refined for internal correctness, faithfulness, or robustness to input variation.

More recently, reasoning-specialized LLMs (e.g., DeepSeek-R1~\cite{deepseekai2025deepseekr1incentivizingreasoningcapability}, OpenAI o4-mini~\cite{openai2024openaio1card}, Gemini 2.5 Flash~\cite{comanici2025gemini}, and Claude 4 Sonnet~\cite{claude4}) have been proposed, which represent a new class of LLMs explicitly optimized for multi-step inference rather than solely for general-purpose language understanding. Unlike conventional instruction-tuned models (e.g., GPT-3.5, LLaMA-2, and Mistral), which are primarily trained on single-turn question–answer or short reasoning tasks, these models incorporate architectural, training, and inference-time modifications targeted at sustaining logical consistency and intermediate-step accuracy. 
For example, OpenAI’s o4-mini~\cite{openai2024openaio1card} introduces extended context windows, optimized attention patterns for long-form reasoning, and inference-time ``deliberation'' passes before producing final answers. 
DeepSeek-R1~\cite{deepseekai2025deepseekr1incentivizingreasoningcapability} adopts a multi-stage training process combining reasoning-focused supervised fine-tuning (SFT) with reinforcement learning from both outcome and process rewards, encouraging faithful and verifiable intermediate steps. 
These reasoning-specialized models also tend to support self-verification loops and iterative refinement during decoding, making them particularly relevant for domains such as program execution inference, where correctness depends on precise, step-by-step reasoning and the handling of edge and invalid inputs.

In this paper, we conduct the first study to systematically evaluate and characterize the reasoning quality of four state-of-the-art reasoning-oriented LLMs, with a particular focus on their failure modes in program execution inference. Our analysis goes beyond measuring final-answer accuracy by examining the correctness, robustness, and common error patterns in the models’ reasoning processes when handling regular, edge-case, and invalid inputs. 
Note that, following existing work~\cite{chen2024reasoning,beger2025coconut,tang2025codereasoner}, and given that the reasoning-specialized models examined in this study already possess built-in multi-step reasoning capabilities, we did not employ Chain-of-Thought (CoT) prompting.

\subsection{Code Reasoning with Large Language Models}
\label{sec:b1}

Code reasoning refers to the ability of a model to understand, analyze, and logically deduce the behavior of programs, often requiring step-by-step inference over control flow, data structures, and state changes~\cite{creswell2022selection}. 
Recently, Gu et al.~\cite{gu2024cruxeval} introduced CRUXEval, which requires LLMs to reason about pre- and post-execution information, such as the given program’s input and output. Building on this idea, Liu et al.~\cite{liu2024codemind} extended the evaluation tasks, specifically, predicting inputs and outputs, to also incorporate natural language specifications. Xue et al.~\cite{xue2025empirical} conducted a 21-participant human evaluation work to evaluate
reasoning quality along with efficiency, logic consistency, and completeness criteria, providing insights into how humans perceive LLM reasoning. 
Zhang et al.~\cite{zhang2025they} empirically explored the external and internal factors of why LLMs generate unsatisfactory CoTs by analyzing 1,023 failed code samples on two
widely used code generation benchmarks. 
Recently, Xu et al.~\cite{xu2024cruxeval} proposed CRUXEVAL-X, a multi-lingual code reasoning benchmark that contains 19 programming languages.  
Dong et al.~\cite{dong2025beyond} evaluated LLMs’ type inference performance on Java code snippets to better understand the strengths and limitations of LLMs on type inference tasks. 
Liu et al.~\cite{liu2025tool} presented ExeRScope, a tool utilizing static and dynamic program analysis to provide additional insights into the code execution reasoning abilities of LLMs. Specifically, it extracted program properties such as programming constructs, program complexity, dynamic program properties, and variable types. It then analyzed LLM’s performance in code execution reasoning concerning these properties. 
Overall, these evaluation methods remain constrained to pre-/post-execution information and do not account for the model’s reasoning about intermediate runtime behavior, such as state changes, control-flow execution paths, and variable updates during program execution. 

Malfa et al.~\cite{la2024code} studied to what extent LLMs can simulate coding and algorithmic tasks to provide insights into general capabilities in such algorithmic reasoning tasks. Specifically, they leveraged the analogy between LLMs and CPUs to study the former as analog simulators of digital devices on algorithms.  
Liu et al.~\cite{liu2024codemind} proposed CodeMind, a framework for evaluating LLMs on code reasoning tasks.  
Chen et al.~\cite{chen2024reasoning} proposed REval, a framework for comprehensively evaluating the code reasoning abilities of code LLMs. They constructed an adapted benchmark based on HumanEval and ClassEval, and developed an evaluation harness to support the framework. Using this setup, they conducted a large-scale empirical study on diverse LLMs, revealing key limitations in reasoning about runtime behavior and information consistency in code models. Claas et al.~\cite{beger2025coconut} proposed CoCoNUT,  a benchmark
for evaluating LLMs' capability to trace the control flow of a
program given relevant input. 

The above studies assessing LLMs’ capability to reason about the runtime behavior of program execution have primarily focused on comparing performance across models. However, they largely treat the reasoning process as a black box, emphasizing output accuracy without examining how the models arrive at their answers. 
In contrast, this paper takes a process-centric perspective: we systematically collect, annotate, and analyze reasoning traces from multiple state-of-the-art reasoning LLMs on program execution inference tasks. We introduce a three-tier input regime (regular, edge-case, invalid) to stress-test models across diverse execution scenarios, develop a taxonomy of reasoning errors at both statement and procedural levels, and identify common failure patterns that degrade execution reasoning performance.
\section{Approach}
\label{sec:approach}

\begin{figure}[t!] 
\centering 
\includegraphics[width=0.85\linewidth]{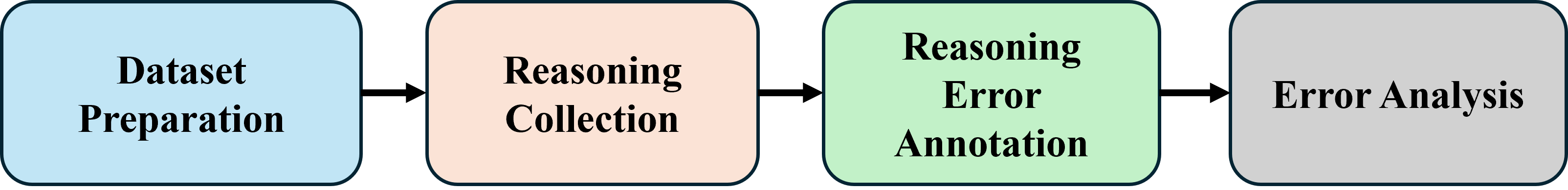} \caption{Overview of our empirical study} 
\label{fig:overview} 
\end{figure}

Figure~\ref{fig:overview} presents an overview of our empirical study, which consists of four steps. Step 1: Dataset Preparation (Section~\ref{sec:3.1}), we construct experimental datasets containing three types of inputs: regular, edge, and invalid. Step 2: Reasoning Collection (Section~\ref{sec:3.2}), we run four state-of-the-art reasoning LLMs to infer the output of a program from our experiment dataset, given a specific input. Step 3: Reasoning Error Annotation (Section~\ref{sec:3.3}), we execute the programs with the given inputs to obtain the ground-truth outputs, which are then used to label the correctness of the LLMs’ inferred results. Step 4: Error Analysis (Section~\ref{sec:3.4}), we analyze the incorrect cases and summarize common reasoning error patterns. 

\subsection{Dataset Preparation}
\label{sec:3.1}

In this section, we first describe the datasets from which the code samples are drawn, ensuring a diverse coverage of programming constructs and difficulty levels. We then outline our input generation methodology, which produces multiple categories of inputs, such as regular, edge case, and invalid examples, to comprehensively evaluate the reasoning capabilities of LLMs under varying conditions. These inputs serve as the basis for assessing each model’s ability to simulate program execution and produce accurate results. 

\subsubsection{Experiment Datasets}

To construct a reliable dataset for evaluating code execution result inference, we reuse two widely adopted code generation benchmarks, i.e., %LeetCode~\cite{xia2025leetcodedataset}, 
HumanEval+~\cite{liu2023your} and LiveCodeBench~\cite{jain2024livecodebench}. 
Note that we use all available data samples from %LeetCode and 
HumanEval+. In contrast, for LiveCodeBench, we selectively include only those tasks labeled as medium or hard. This selection criterion is motivated by the need to focus our evaluation on code problems that present a higher level of reasoning and computational challenge. Easy-level tasks in LiveCodeBench often involve straightforward syntax manipulations or simple logic that may not sufficiently stress the multi-step reasoning capabilities of large language models. 
Overall, these tasks cover a diverse range of programming problems at varying difficulty levels, each accompanied by corresponding test cases and reference implementations. We use the ground-truth solutions provided by each benchmark as the target programs in our experiments. For each dataset, we report the number of tasks, the average lines of code (LOC), and code complexity metrics, i.e., Cyclomatic Complexity~\cite{ebert2016cyclomatic} and Halstead Metrics~\cite{hariprasad2017software}. Detailed statistics are presented in Table~\ref{tab:dataintro}. Overall, the code complexity metrics indicate that LiveCodeBench contains the most challenging programs, %followed by LiveCodeBench, 
while HumanEval+ comprises less complex programs.

%To ensure high-quality and realistic code examples, we collect the generated correct solutions (i.e., passing the given test cases) from top-performing models listed on each benchmark’s leaderboard. \mamad{We used the solutions reported by the datasets}By using outputs from state-of-the-art models (e.g., GPT-4, Claude, CodeLLaMA), we focus our study on code that is syntactically valid, semantically rich, and representative of what modern LLMs produce in real-world settings. These collected programs serve as the subject programs for evaluating whether LLMs can accurately infer their execution results without actually running them.

%livecodebench

%humaneval

\subsubsection{Input Generation}

For each program, we evaluate the LLMs’ ability to handle regular inputs, edge case inputs, and invalid inputs. Our strategy for the input generation is as follows. %\todo{Mohammad: update the strategies you used, if the desc are not accurate}
\begin{itemize}

  \item \textbf{Regular Inputs}: These inputs are directly taken from the original test cases provided by the benchmark datasets. They represent typical, well-formed inputs that meet all problem constraints. If a problem contains multiple official test cases, we randomly sample or use all of them to ensure sufficient coverage of standard execution scenarios. It should be noted that we applied an Input Space Partitioning (ISP)\cite{InputSpacePartioning} approach to guide the selection process. Specifically, we identified the relevant input parameters and characteristics for each problem, partitioned the input domain into disjoint and complete blocks, and then selected representative values from each block. This ensured that the chosen inputs collectively cover all identified partitions and capture diverse representative behaviors within the valid input space.

  \item \textbf{Edge-Case Inputs}: These are inputs that lie at or near the boundaries of the problem constraints. We construct them by systematically varying parameters to their minimum and maximum allowed values, introducing zero values where applicable, and considering degenerate cases (e.g., empty strings, arrays of size 1, maximum integer values). The goal is to test the LLMs’ ability to reason correctly under extreme but valid conditions.

  \item \textbf{Invalid Inputs}: These inputs were constructed manually to intentionally violate the constraints of the problem and to evaluate the robustness of the LLMs when faced with erroneous or unexpected data. The categories of invalid inputs include: wrong data types (e.g., passing a string where an integer is expected), too many arguments, out-of-range values, missing required arguments, unexpected keyword arguments, division-by-zero inputs, use of \texttt{None} or null values where disallowed, and structurally malformed inputs (e.g., missing required fields). To confirm that these examples were indeed invalid, we executed each case against the reference implementation and verified that they consistently raised errors or exceptions at runtime.  For this type of input, we expect LLMs to recognize it as invalid, rather than to generate any outputs.
\end{itemize}
\begin{figure}[t!]
\centering 
\centering 
\includegraphics[width=0.85\linewidth]{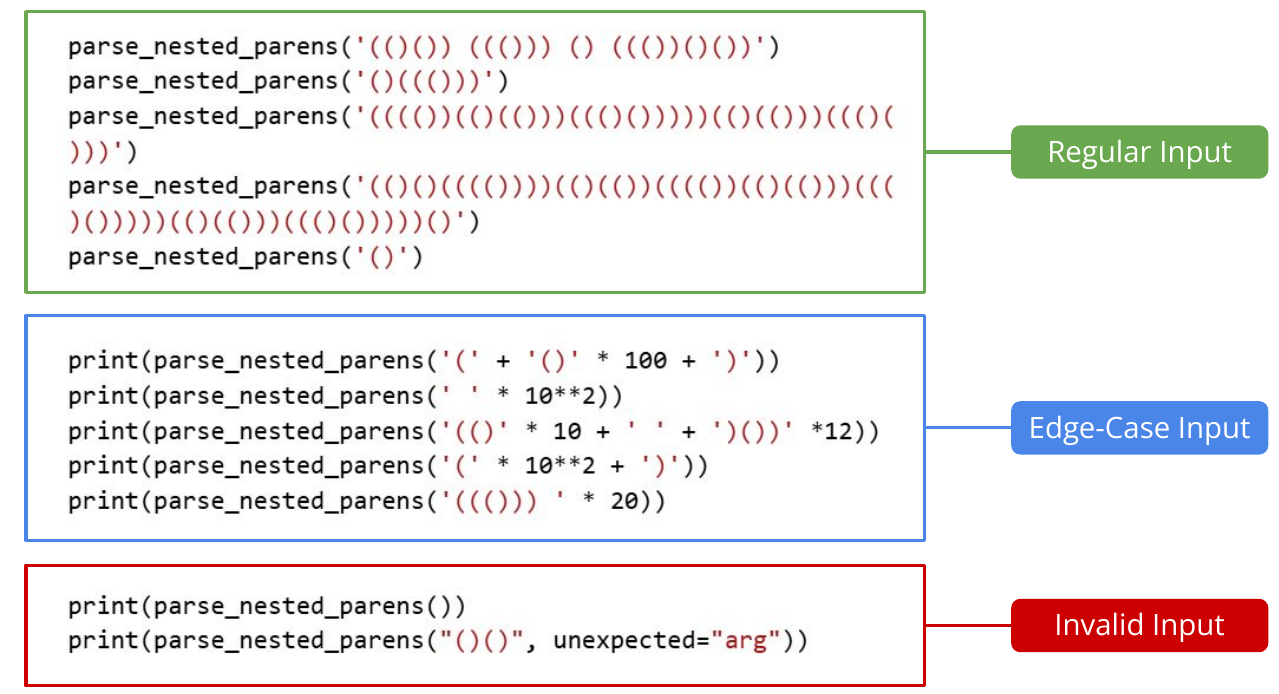} \caption{Three input types for the example in Fig.~\ref{fig:llm_reasoning_output_sample}.} 
\label{fig:input_types} 
\end{figure}

This three-tier input generation strategy enables our evaluation to encompass a broad spectrum of execution scenarios, ranging from normal operation to boundary conditions and invalid cases, thereby providing a comprehensive assessment of LLM reasoning capabilities. Figure~\ref{fig:input_types} shows an example of the three types of inputs for the example program shown in Figure~\ref{fig:llm_reasoning_output_sample}. 

For each program in our study (HumanEval+~\cite{liu2023your} and LiveCodeBench~\cite{jain2024livecodebench}), we generate five regular inputs, five edge-case inputs, and two invalid inputs. In total, this results in 
$(164+263) * 12=5,124$ inferences per LLM. Considering all four LLMs evaluated, the study involves 20,496 inferences generated by LLMs.

% \song{taking the example you showed in Intro, to list the three types of inputs used for this example}

\begin{table}[t!]
\centering
\caption{Overview of the datasets. \textbf{CC} denotes the Cyclomatic Complexity metric, and \textbf{HM} is the Halstead Metrics.}
\label{tab:dataintro}

\begin{tabular}{|l|c|c|c|c|}
\hline
\textbf{Dataset}  & \textbf{\#Task} & \textbf{Avg. \#LOC} & \textbf{Avg. CC}  & \textbf{Avg. HM} \\ \hline
%LeetCode      &    228    &    29   &  10.7  &   4.2  \\ \hline
HumanEval+   &    164    &    8   &  3.1  &  2.2   \\ \hline
LiveCodeBench &    263    &    12   &  5  &  3   \\ \hline
\end{tabular}

\end{table}

\subsection{Reasoning Collection}
\label{sec:3.2}

\subsubsection{Reasoning LLM Selection}

In this work, we evaluate four state-of-the-art reasoning-oriented LLMs, i.e., DeepSeek-R1~\cite{deepseekai2025deepseekr1incentivizingreasoningcapability}, OpenAI o4-mini~\cite{openai2024openaio1card}, Gemini 2.5 Flash~\cite{comanici2025gemini}, and Claude 4 Sonnet~\cite{claude4}, on their ability to infer program execution results given three categories of inputs: regular, edge-case, and invalid. 
These models were selected for the following reasons: (1) they represent the latest generation of reasoning-focused architectures from leading AI providers; (2) they have demonstrated strong performance in complex reasoning benchmarks, including mathematical problem-solving, code generation, and multi-step logical inference~\cite{abdollahi2025surveying}; and (3) they cover a diverse range of model architectures and training paradigms, enabling a balanced and representative comparison across different reasoning approaches. 

It is worth noting that prior work~\cite{chen2024reasoning} primarily employed LLMs with relatively small parameter sizes that are not explicitly designed or optimized for advanced reasoning tasks, for example, CodeLlama (7B–34B), Magicoder-7B, StarCoder2 (3B–15B), Mistral, Gemma, and GPT-3.5. In contrast, our study intentionally excludes such models, as they do not fall into the category of reasoning-specialized LLMs and typically lack the multi-step inference capabilities we aim to evaluate.

\subsubsection{Process of Reasoning Collection}

For each program and its corresponding input (i.e., program-input pair), we prompt all four reasoning-oriented LLMs to generate an explicit thinking process that traces how the execution output is derived (Detailed prompts are available in our replication package). 
Rather than focusing solely on the final prediction, this setup compels the models to articulate intermediate reasoning steps that reflect runtime behavior. 
To further strengthen performance and semantic fidelity, we also instruct the LLMs to refine their code reasoning while explicitly considering control flow (e.g., branching, loops, and execution order) and data flow (e.g., variable dependencies and state transitions), which grounds the reasoning process in program semantics, enabling a more faithful alignment between the models’ inferred reasoning traces and the actual runtime execution behavior. 
Note that, for the invalid inputs, we expect LLMs to recognize them as invalid, rather than to generate any outputs. 

\subsection{Reasoning Error Annotation}
\label{sec:3.3}
After prompting the LLMs to generate reasoning traces for each program–input pair, we collected all samples in which the models produced incorrect outputs or could not recognize the invalid inputs. Since the final outputs are directly derived from the reasoning traces, we hypothesized that errors in the outputs likely reflect flaws in the underlying reasoning.  
%To investigate this connection, we conducted a detailed manual analysis of the erroneous reasoning traces to uncover potential root causes and recurring error patterns.

At this stage, three authors independently labeled the data, identifying both where in the program execution the reasoning first failed and why the reasoning as a whole went wrong. 
For each program–input pair where the LLM produced an incorrect output, we extracted the corresponding reasoning trace and treated it as an annotation unit. Each failed reasoning trace was examined at two levels:
\begin{itemize}

\item \textbf{Statement-level failures:} These capture the precise execution point where the model’s reasoning first deviates from the ground-truth program behavior. Instead of examining the trace as a whole, statement-level analysis focuses on the specific program statement (e.g., a conditional branch, loop iteration, function call, or variable assignment) where the reasoning went astray. For example, the model may misinterpret the outcome of a conditional, apply the wrong update to a variable inside a loop, or fail to expand a recursive call correctly. By pinpointing these divergence points, statement-level annotations reveal the immediate source of error that propagates through the remainder of the reasoning trace.

\item \textbf{Trace-level failures:} these refer to the broader reasoning breakdowns that characterize an entire incorrect reasoning trace, beyond a single misstep. While statement-level failures capture where the model first deviates from the correct execution, trace-level failures summarize why the reasoning as a whole went wrong. For example, a trace-level failure may involve logical contradictions across different steps in the trace, repeated propagation of an earlier miscalculation, or hallucinated steps that introduce behavior absent from the source code.

\end{itemize}
The independently generated labels were then reviewed collaboratively by three extra authors, and any disagreements were resolved through consensus. 
This process yielded a comprehensive set of annotated reasoning errors that captured both the local points of failure, i.e., the exact statements where reasoning first diverged, and the trace-level failures that characterized each incorrect reasoning trace as a whole.

% \song{taking the example you showed in Intro, to list the two types of annotation info for this example}
\begin{wrapfigure}{r}{0.5\textwidth}
\centering 
\includegraphics[width=1\linewidth]{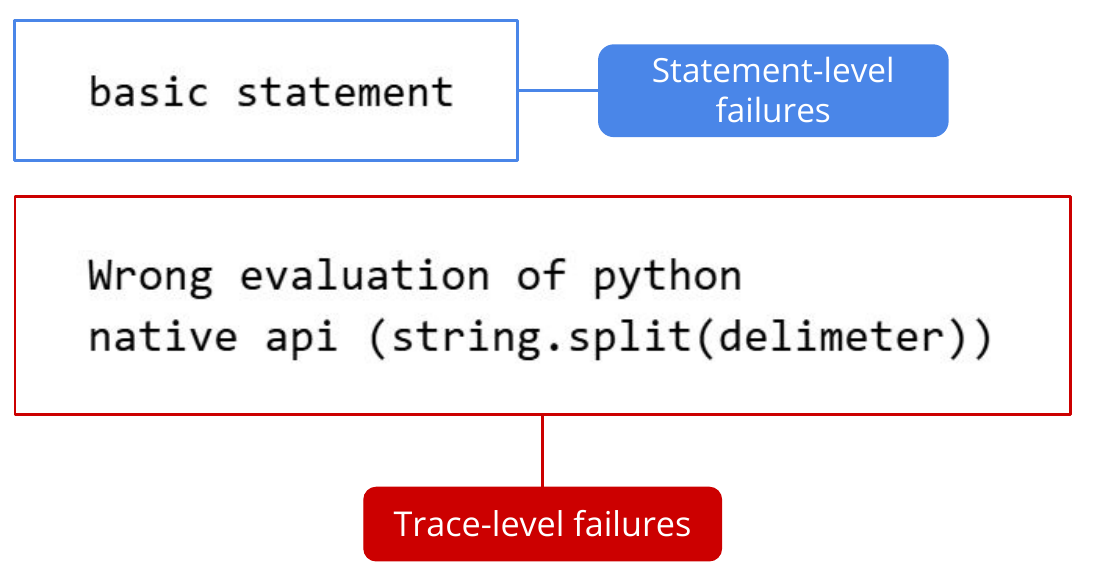} \caption{Reasoning error annotation for the reasoning error in Fig.~\ref{fig:llm_reasoning_output_sample}} 
\label{fig:error_analysis} 
\end{wrapfigure}

\subsection{Error Analysis}
\label{sec:3.4}

In this stage, we leveraged the annotated data from the previous section to uncover potential root causes and recurring error patterns that occurred across common failure cases. To ensure rigor and minimize individual bias, we adopted an open card sorting approach, a qualitative analysis technique widely used to surface latent categories from unstructured data. Each failed reasoning trace was treated as an individual ``card'', and annotators grouped similar cases to form preliminary categories.

Specifically, we focused on reviewing the trace-level failure reasons present in these data. Each failed reasoning trace was carefully examined to identify shared characteristics, recurring patterns, or common failure modes. This collaborative process encouraged discussion and cross-validation among all six authors, helping to reduce subjective bias and improve the reliability of the categorizations. 
%Following the independent labeling, the authors engaged in collaborative review sessions. Any disagreements in annotations were resolved through discussion until consensus was reached, ensuring that the final categorization reflected a shared understanding rather than individual judgment. Because disagreements were resolved by consensus, the calculation of inter-rater agreement (e.g., Cohen’s k) was not applicable. Instead, transparency was maintained through detailed documentation of the decision-making process and the rationale for merging or splitting categories.
Finally, after consensus was reached on the initial groupings, all authors collaboratively refined and consolidated these categories into a coherent taxonomy of reasoning error types. %This taxonomy captures both high-level patterns (e.g., control-flow misalignment, data-flow inconsistency \song{update here when you have concrete patterns}) and fine-grained subtypes (e.g., mishandling of nested conditionals, overwriting of state variables, or premature termination of reasoning \song{update here when you have concrete patterns}). 
This iterative process not only yielded a structured framework for characterizing reasoning failures but also provided a foundation for subsequent quantitative analyses of error distribution and frequency. For example, Figure~\ref{fig:error_analysis} presents an illustrative example of our error analysis. By examining the code snippet and the corresponding generated reasoning shown in Figure~\ref{fig:llm_reasoning_output_sample}, we identified that the reasoning failure happens in a basic method call statement and stems from the incorrect use of the Python split function. %According to our taxonomy, this case is categorized as a wrong evaluation for Python native API at the trace-level, and as a basic statement error at the statement-level.

%\begin{itemize}
 
%\item \textbf{Step 1}: collect the  problem description, reference solutions for tasks in humaneval(164 test samples：\url{https://huggingface.co/datasets/openai/openai_humaneval}) and livecodebench (479 test samples: \url{https://huggingface.co/datasets/livecodebench/execution-v2}) and LeetCodeDataset (228 test samples): \url{https://huggingface.co/datasets/newfacade/LeetCodeDataset/viewer/default/test}

% I should mention the leetcode dataset too, also I should provide analysis of code complexity metrics for each dataset (like different charts for each dataset showing different complexity metrics) that proves our datasets were through and have included a wide range of complexities. 

%\item \textbf{Step 2}: Run the LLM Think models on the executable code of test samples and collect 1) the subset of issues that each model can resolve, 2) the left issues that each model cannot resolve (we will explore whether LLMs show understanding capability on the issues they can resolve and cannot resolve)

%\item \textbf{Step 3:} For each issue collected from Step 2, randomly generate inputs covering edge cases, regular inputs, and illegal inputs. Run the solutions with these inputs and collect the ground-truth outputs.

%\item \textbf{Step 4:} Given the solution (option2: solution + task description) to LLMs and ask them to produce step-by-step thinking and infer the results.

%\item \textbf{Step 5:} Manually check the failed cases where LLMs cannot infer correct outputs and summarize the reasons. Exploring whether LLMs show understanding capability on the issues they can resolve and cannot resolve.

%\end{itemize}

% \usepackage{multirow}
\begin{table}[t!]
\centering
\caption{The number of unsuccessful inferences (i.e., program–input pairs) for the four reasoning models across different input types. Numbers in parentheses denote the count of unique programs within those unsuccessful cases.} 
\label{tab:unsuccessful}
\setlength{\tabcolsep}{2pt}
\scalebox{0.9}{
\begin{tabular}{|c|ll|cc|cc|}
\hline
\multirow{2}{*}{Models} & \multicolumn{2}{c|}{regular inputs}& \multicolumn{2}{c|}{edge inputs} & \multicolumn{2}{c|}{invalid inputs} \\ \cline{2-7} 
& \multicolumn{1}{l|}{HumanEval+} & LiveCodeBench &  \multicolumn{1}{c|}{HumanEval+} & LiveCodeBench & \multicolumn{1}{c|}{HumanEval+} & LiveCodeBench \\ \hline
Gemini & \multicolumn{1}{c|}{44 (26)}& 77 (45) & \multicolumn{1}{c|}{77 (56)}& 113 (84) & \multicolumn{1}{c|}{25 (23)} & 68 (64)\\ \hline
GPT-4o&  \multicolumn{1}{c|}{25 (15)}& 25 (20) & \multicolumn{1}{c|}{84 (56)}& 107 (89) & \multicolumn{1}{c|}{6 (6)} & 48 (44)\\ \hline
Deepseek R1 & \multicolumn{1}{c|}{39 (21)}& 118 (62) & \multicolumn{1}{c|}{103 (65)}& 175 (125) & \multicolumn{1}{c|}{7 (7)} & 38 (36)  \\ \hline
Claude 4& \multicolumn{1}{c|}{48 (26)}& 127 (74) & \multicolumn{1}{c|}{84 (52)}& 178 (126) & \multicolumn{1}{c|}{24 (23)} & 76 (71)\\ \hline
\end{tabular}
}
\end{table}
\section{Result Analysis}
\label{sec:results}

\subsection{RQ1: Performance of Reasoning LLMs}
\label{rq1}

\begin{figure}[t!] 
\centering 
\includegraphics[width=1\linewidth]{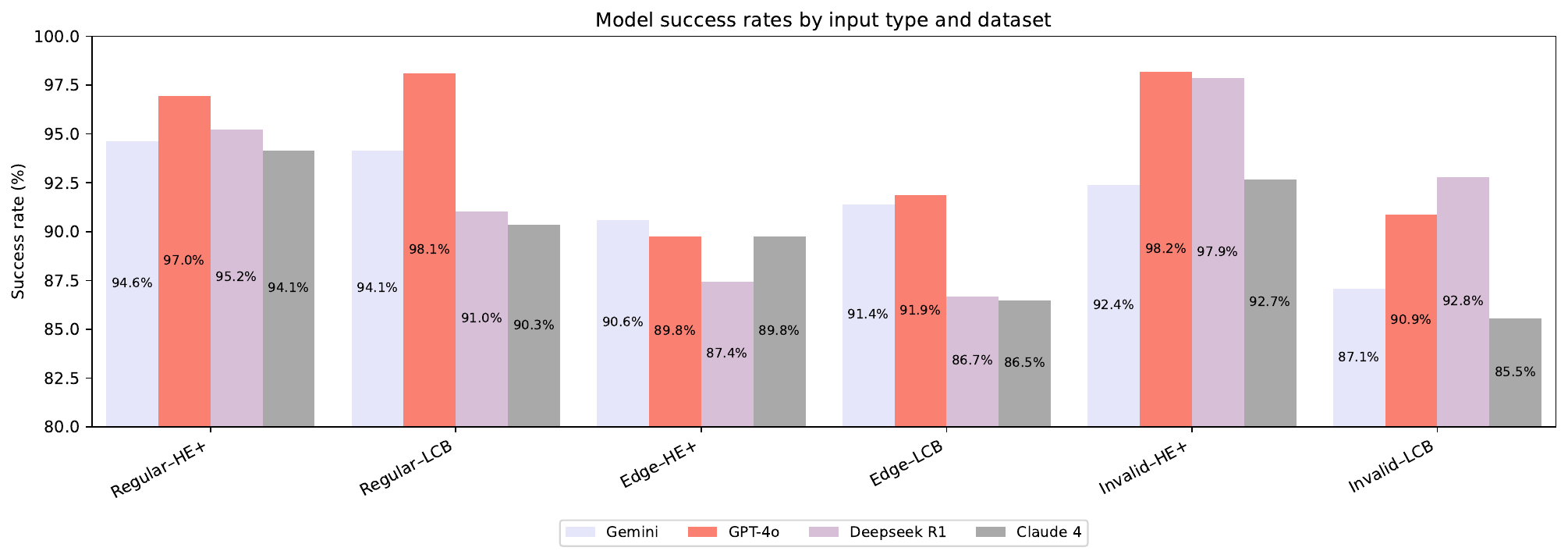} \caption{The success rates of each model across two datasets (HumanEval+ and LiveCodeBench) under different input types (regular, edge, and invalid).} 
\label{fig:ss} 
\end{figure}

%\song{draw a figure to show the success rate of each model on the two datasets for each type; for Huameval+, you have 164*5 (820) program-input pairs for regular/edge and 164*2 program-input pairs for invalid inputs, show the rate for each model; for LiveCodeBench, just do the same}

%\song{@Rifa: you can calculate the numbers from table 2, which shows the failed program-input pairs}

%\song{describe your figure here}

Figure~\ref{fig:ss} presents the success rates of the four experimented reasoning LLMs, i.e., Gemini, GPT-4o, Deepseek R1, and Claude 4, on \textit{HumanEval+} and \textit{LiveCodeBench}, evaluated across three input types, i.e., regular, edge, and invalid inputs. Among the models, GPT-4o shows the most consistent and robust performance, achieving the highest accuracy for regular inputs with 97\% on HumanEval+ and 98.1\% on LiveCodeBench. Deepseek R1 and Gemini also perform well in this category, while Claude 4 trails slightly behind.  
The performance gap widens when models are evaluated on edge and invalid inputs. For edge inputs, Gemini achieves the highest success rate on HumanEval+ (90.6\%), while GPT-4o leads on LiveCodeBench (91.9\%). On invalid inputs, GPT-4o achieves the top score on HumanEval+ (98.2\%), whereas Deepseek R1 performs best on LiveCodeBench (92.8\%). Claude 4 consistently shows lower success rates in these challenging scenarios, particularly on LiveCodeBench, where its performance drops to 85.5\% for invalid inputs. These results indicate that while all models are reliable on standard inputs, GPT-4o and Deepseek R1 in particular maintain high performance under more difficult or adversarial conditions.

We also show the detailed number of unsuccessful inferences (i.e., program-input pairs) in Table \ref{tab:unsuccessful}. 
% \song{describe Table2} 
%Table \ref{tab:unsuccessful} reports the detailed number of unsuccessful inferences (i.e., program-input pairs) for the four reasoning LLMs across the three different types of inputs on the two widely used benchmarks, HumanEval+ (164 programs) and LiveCodeBench (263 programs).  
%The numbers outside parentheses indicate the total number of failures, while the numbers in parentheses represent the count of unique programs for which models failed. For example, on the HumanEval+ dataset, Gemini failed to produce correct results for 44 program–input pairs in the regular input category, corresponding to 26 unique programs. 
From the table, we observe that all models consistently exhibit higher numbers of unsuccessful inferences on LiveCodeBench compared to HumanEval+. This gap can be attributed to the inherent difficulty of the tasks, as programs in LiveCodeBench are more challenging than those in HumanEval+ (see Table~\ref{tab:unsuccessful}). 
We can observe that all models fail roughly twice as often on edge inputs compared to regular inputs. This suggests that while the models are generally capable of handling typical cases, they struggle when confronted with boundary conditions or inputs that push the limits of the problem constraints.  
Gemini and Claude 4 perform consistently across both valid (regular and edge) and invalid inputs, demonstrating stable behavior regardless of input type. In contrast, GPT-4o and Deepseek R1 perform significantly better at identifying invalid inputs, which could be attributed to differences in their training objectives or input validation mechanisms. 

\mybox{\textbf{Answer to RQ1:} The examined reasoning LLMs achieve high success rates across all three input types on both HumanEval+ and LiveCodeBench, consistently above 85\% and can be up to 98.17\%.  Overall, all models perform better on regular and invalid inputs than on edge cases, likely due to their higher complexity and limited representation in the training data.}

\subsection{RQ2: Failure Locations}
\label{rq2}
% \begin{figure}[t!] 
% \centering 
% \includegraphics[width=0.9\linewidth]{figs/error_locations1.pdf} \caption{Error locations\song{use latex to draw this, check the source of Nima's paper I shared with you months before}} 
% \label{fig:error_locations} 
% \end{figure}

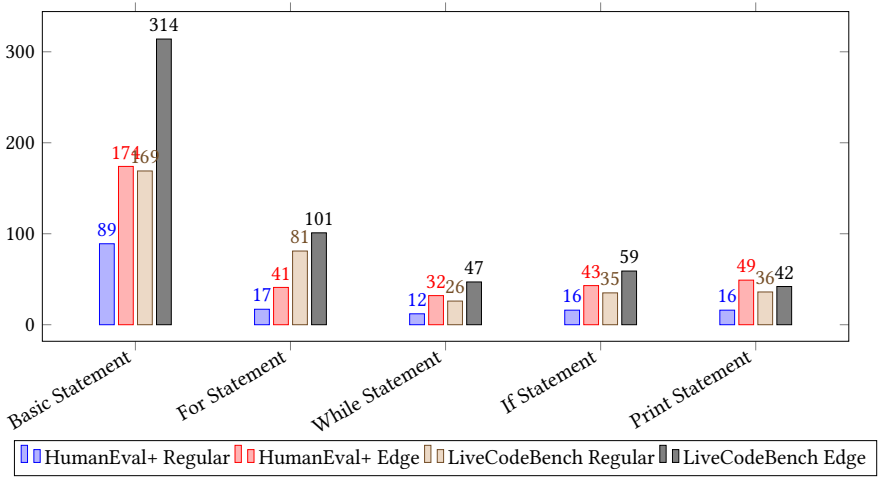
\begin{figure}[t!]
\centering
\begin{tikzpicture}[scale=0.8]
\begin{axis}[
    ybar,
    bar width=7pt,
    width=15cm,
    height=7cm,
    ylabel={},
    xlabel={},
    symbolic x coords={Basic Statement, For Statement, While Statement, If Statement, Print Statement},
    xtick=data,
    nodes near coords,
    nodes near coords align={vertical},
    xticklabel style={rotate=30, anchor=east},
    legend style={at={(0.5,-0.3)}, anchor=north,legend columns=-2},
    enlarge x limits=0.15
]

% HumanEval+ Regular Inputs
\addplot coordinates {(Basic Statement,89) (For Statement,17) (While Statement,12) (If Statement,16) (Print Statement,16)};
% HumanEval+ Edge Inputs
\addplot coordinates {(Basic Statement,174) (For Statement,41) (While Statement,32) (If Statement,43) (Print Statement,49)};
% LiveCodeBench Regular Inputs
\addplot coordinates {(Basic Statement,169) (For Statement,81) (While Statement,26) (If Statement,35) (Print Statement,36)};
% LiveCodeBench Edge Inputs
\addplot coordinates {(Basic Statement,314) (For Statement,101) (While Statement,47) (If Statement,59) (Print Statement,42)};

\legend{HumanEval+ Regular, HumanEval+ Edge, LiveCodeBench Regular, LiveCodeBench Edge}
\end{axis}
\end{tikzpicture}
\caption{Distribution of locations of reasoning failures across all unsuccessful inferences for all LLMs.}
\label{fig:error_locations}
\end{figure}

In this research question, we investigate which statements or code locations most frequently trigger errors in LLMs’ reasoning during code inference. 
Figure \ref{fig:error_locations} summarizes the distribution of reasoning failures' locations in model-generated reasoning traces across both datasets. Since this location information does not apply to invalid inputs, the analysis includes only regular and edge inputs, highlighting the program structures where failures occur most frequently. 

As we can see from the Figure, basic statements dominate the failure locations. These include operations such as variable assignments, arithmetic updates, value substitutions, etc. In HumanEval+, basic statements account for 89 failures under regular input and 174 under edge input, corresponding to more than half of the total errors. This pattern is amplified in LiveCodeBench, where 169 and 314 failures are linked to basic statements in the regular and edge input categories, respectively. These findings suggest that even straightforward computational steps (e.g., arithmetic updates, variable assignments) remain a major bottleneck for current reasoning models, particularly when embedded in more complex code contexts. 

Control flow structures (loops and conditionals) also emerge as critical sources of difficulty. The \textit{For Statements} show a dramatic rise in errors between HumanEval+ and LiveCodeBench regular inputs and edge inputs. This reflects the fact that LiveCodeBench includes more algorithmically challenging problems where iteration and index management are central. The \textit{While Statements} and \textit{If Statements} similarly exhibit growth, though less pronounced. In particular, inference failures stem from the \textit{If} conditions nearly triple from HumanEval+ regular inputs to LiveCodeBench regular inputs, highlighting the models' struggles with branching logic when problem specifications are more intricate.

The unexpected result is the error profile of the \textit{Print Statements}. While these are not algorithmically complex structures, they produce a non-negligible share of failures, i.e., 16 on HumanEval+ regular inputs and 36 on LiveCodeBench regular inputs. Under edge inputs, print-related failures rise to 49 and 42, respectively. This indicates that formatting, output handling, or misreporting of final results are persistent weak points, especially when edge cases stress correct reporting of intermediate or final values. 
%
%(1) \textbf{Basic statements} (e.g., assignments, arithmetic updates, substitutions) dominate the error space in both datasets, accounting for more than half of all failures. \\ (2) \textbf{Control-flow structures} (loops and conditionals) are another significant source of errors, with for-loop failures rising sharply from HumanEval+ to LiveCodeBench, and if-statement errors nearly tripling. \\ (3) These patterns confirm that iteration, branching, and index management are challenging for LLMs, especially with more algorithmically complex tasks. \\ (4) \textbf{Print statements}, despite being simple constructs, show a surprisingly high error rate, highlighting persistent weaknesses in the output formatting and reporting of the results, particularly under edge inputs.
%
\mybox{\textbf{Answer to RQ2:} Basic statements like assignments and arithmetic dominate reasoning failures, while control flow structures, loops, and conditionals also contribute significantly, especially in complex tasks. For-loop and if-statement errors rise sharply from HumanEval+ to LiveCodeBench. Even simple print statements show high error rates, revealing persistent weaknesses in output formatting and edge-case handling.}

\subsection{RQ3: Failure Patterns}
\label{rq3}
\begin{figure}[t!] 
\centering 
\includegraphics[width=0.95\linewidth]{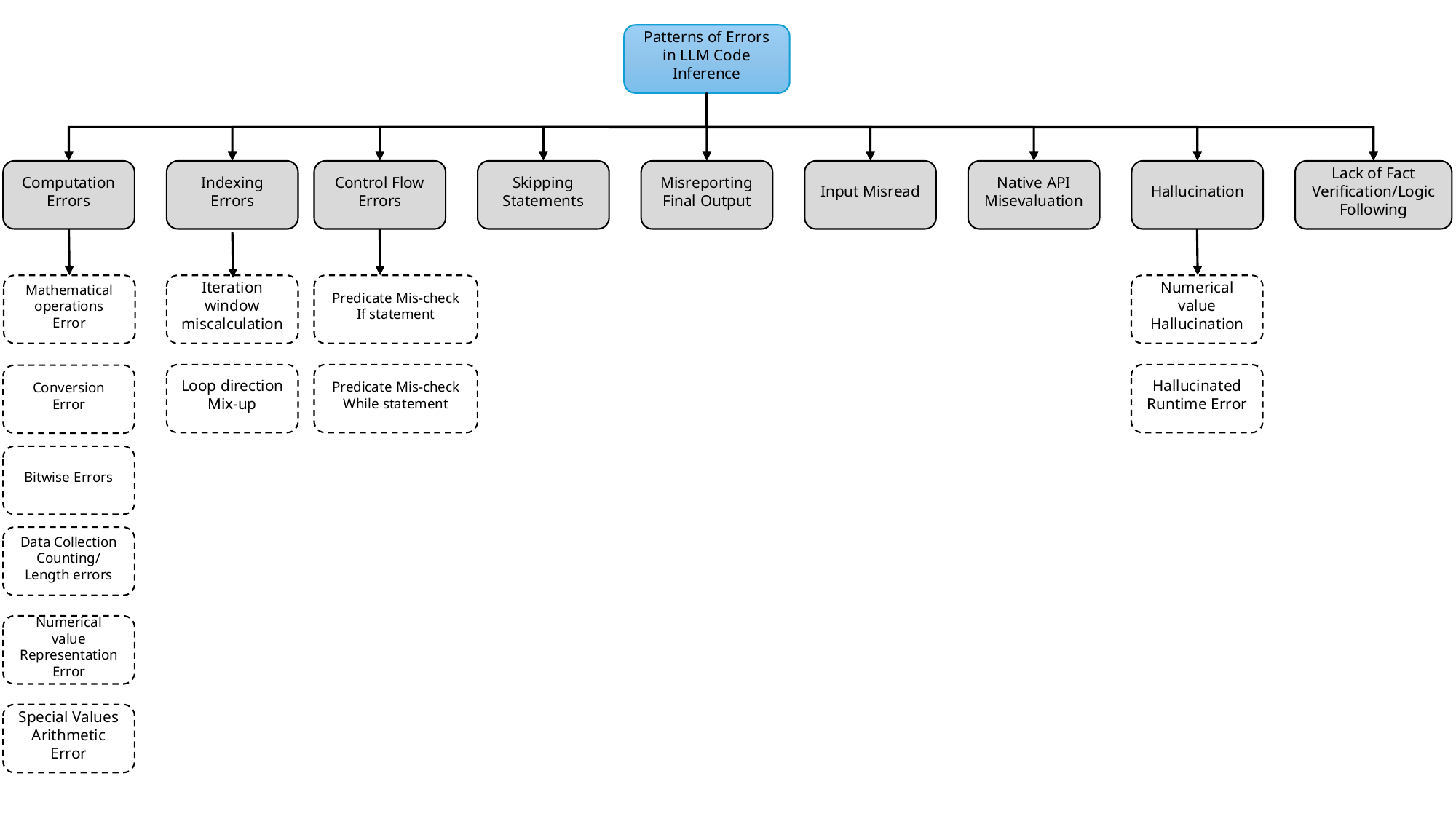} 
\vspace{-0.1in}
\caption{Error Taxonomy of Reasoning Failures in LLM-Based Code Inference} 
\label{fig:error_taxonomy} 
\end{figure}

Based on the error analysis process described in Section~\ref{sec:3.4}, we derived a taxonomy that categorizes the main types of inference failures exhibited by LLM during code inference. Figure~\ref{fig:error_taxonomy} provides a visual overview of this taxonomy. The goal of this classification is to structure the diverse failure modes observed in our experiments into well-defined categories and subcategories, enabling systematic interpretation and comparison between datasets and models.  

% \song{@M： overall, the explanation of each type is too simple; give more details; for each category or sub-category, give a simplified concrete example; for the example, you can refer to  the problem ID in each dataset and describe the issues in the reasoning generated for the example}

The taxonomy comprises nine high-level categories, some of which are further refined into specific subcategories. The categories include: 

\textbf{1. Computation Errors.}  
This category covers issues in arithmetic, type conversions, number representation, and related operations.
\begin{itemize}
    \item \textit{Mathematical Operations Error}: Failures in basic arithmetic operations, such as addition, subtraction, multiplication, or division. 
    
    \textit{Example\footnote{Details of the generated reasoning traces are provided in the replication package; due to space constraints, they are not included in this paper.}:} Gemini produced incorrect results for the task \textit{HumanEval-130}, with the failure originating from an arithmetic miscalculation at iteration $i = 3$. Specifically, the expression $2 + 3 + 1 + (4/2)$ was computed as 9 instead of the correct value 8. This error then propagated forward, leading to incorrect results for subsequent terms at odd indices. 
  
    \item \textit{Conversion Error}: Missteps in type conversions, for example, incorrect handling of \texttt{int $\rightarrow$ float}, \texttt{string $\rightarrow$ int}, or \texttt{int $\rightarrow$ bits}. 
    
    \textit{Example:} Claude generated incorrect output for the task \textit{LiveCodeBench-80}, with the error originating from an incorrect integer-to-binary conversion when evaluating \\ \texttt{bin(1000000000000000000)}.
    
    \item \textit{Bitwise Errors}: Incorrect evaluations of bitwise operations such as \texttt{xor}, \texttt{and}, or \texttt{shift}.
    
    \textit{Example:} Gemini produced incorrect results for the task \textit{LiveCodeBench-14}, with the failure caused by an incorrect XOR operation $(2^{64} - 1) \oplus (-2^{63})$.
    
    \item \textit{Data Collection Counting/Length Errors}: Miscounting the length of arrays or collections, often leading to off-by-one errors. 
    
    \textit{Example:} Gemini produced incorrect results for the task \textit{HumanEval-111}. The failure originated from a miscount of the collection elements. The input list contained 100 elements (50 `a` and 50 `b`), but the count was incorrectly reported as $48 + 48 = 96$ elements.

    \item \textit{Numerical Value Representation Error}: Issues with floating-point representation, rounding, or numerical stability. 
    
    \textit{Example:} Gemini generated inaccurate results for the task \textit{HumanEval-2}, where the value 72999331.86348532 was stored as 72999331.8634853214025497, indicating its floating-point precision limitations.
    
    \item \textit{Special Values Arithmetic Error}:  Mishandling of special values such as \texttt{NaN} or \texttt{Inf}.  
    
    \textit{Example:} Gemini failed to produce the correct results for the task \textit{HumanEval-21}, where the case $x = float('-inf + inf')$ was mishandled, with the expression $-inf + inf$ evaluated as 0 instead of the correct result $NaN$.  
\end{itemize}

\vspace{2pt}
\textbf{2. Indexing Errors.}  
These errors occur when accessing or slicing arrays, lists, or strings.  
\begin{itemize}
    \item \textit{Iteration Window Miscalculation}: Selecting an incorrect slice or sub-window of a sequence. 
    
   \textit{Example:} Gemini produced incorrect results for the task \textit{LiveCodeBench-36}, where for a given array $s$, Gemini cannot iterate the elements correctly at the 14th element, and incorrect treat $s[13]$ and $s[14]$ as the same, leading to a missed transition.

    \item \textit{Loop Direction Mix-up}: Iterating in the wrong direction (e.g., forward instead of backward), leading to incomplete or incorrect traversals. 
    
    \textit{Example:} Claude failed to generate the correct results for the task \textit{HumanEval-59}, where the loop \texttt{for i in range (-11, 0, -1)} was misinterpreted as producing the sequence \{$-11, -10, \ldots, -1$\}. In reality, as the step is $-1$ and the stop value $0$ is greater than the start value $-11$, no values are generated and the loop body never executes.
 
\end{itemize}

\vspace{2pt}
\textbf{3. Control Flow Errors.}  
This category captures mistakes in conditional and iterative constructs.  
\begin{itemize}
    \item \textit{Predicate Mischeck (if statement)}: Incorrect evaluation of the truth value of an \texttt{if} condition. 
    
    \textit{Example:} Gemini produced incorrect results for the task \textit{LiveCodeBench-332}, with the failure arising from incorrectly logging \texttt{s[9] == s[8]} as \texttt{True}.
  
    \item \textit{Predicate Mischeck (loop statement)}: Errors in evaluating loop continuation conditions, leading to premature termination or infinite loops. 
    
    \textit{Example:} DeepSeek produced incorrect results for the task \textit{LiveCodeBench-75}, where the loop condition was mis-evaluated. It incorrectly advanced to an iteration it considered valid, leading to an incorrect return value. 
    
\end{itemize}

\vspace{2pt}
\textbf{4. Skipping Statements.} This category covers errors related to omitting necessary logical steps, often resulting in premature jumps to a final result.

\textit{Example:} Gemini produced incorrect results for the task \textit{LiveCodeBench-76}, caused by skipped value-update statements within a loop, which led to an incorrect final result.

\vspace{2pt}
\textbf{5. Misreporting Final Output.}    
In some cases, the model correctly follows the internal reasoning steps, but reports an incorrect final result. This highlights a disconnect between internal reasoning fidelity and external output generation. 

\textit{Example:} Gemini failed to generate accurate results for the task \textit{LiveCodeBench-54}. The model carried out the intermediate steps accurately and reached the correct output \texttt{``aaayyyaba''}. However, while joining the characters, the value was misreported as \texttt{``aaayyyaab''}, indicating that the mistake occurred in the final output construction.  

\vspace{2pt}
\textbf{6. Input Misread.}  
These errors arise when the model misinterprets the input of the problem. Typical cases include reading strings incorrectly, confusing the input length, or overlooking input constraints. 

\textit{Example:} Gemini produced incorrect results for the task \textit{LiveCodeBench-289} due to misreading the input length. The input string was \textit{s = ``00110101101100101011001101''} with a size of 28, but Gemini incorrectly computed the length as 26.

\vspace{2pt}
\textbf{7. Native API Misevaluation.}  
Mistakes in applying Python’s built-in functions or standard libraries fall into this category. Examples include incorrect assumptions about function defaults (e.g., sorting order) or confusing similar functions. 

\textit{Example:} Claude generated incorrect results for the task \textit{HumanEval-162}. The error occurred because the program used \textit{hashlib.md5()} to hash an input, which Claude could not correctly evaluate. As a result, it produced an incorrect hash type instead of the intended MD5 value.

\vspace{2pt}
\textbf{8. Hallucination.}  
Hallucinations occur when the model produces results disconnected from the actual execution of the code.  
\begin{itemize}
    \item \textit{Numerical Value Hallucination}: Fabricating results for very large or complex numbers instead of computing them. 
    
    \textit{Example:} Claude generated inaccurate results for the task \textit{HumanEval-106}. While its reasoning about the rules was mostly correct (i.e., building a list based on different rules for odd and even indices), the even terms were accurately calculated only up to $i = 28$. Beyond that, it hallucinated large numbers instead of computing them, preventing the complete list from being generated.

    \item \textit{Hallucinated Runtime Error}: Incorrectly claiming that the code execution results in an error when it does not. 
    
    \textit{Example:} Gemini produced incorrect results for the task  \textit{HumanEval-36}, the output was assumed to be a ``Time Limit Exceeded'' error. In reality, the program may take longer to run, but ultimately produces the correct output.
   
\end{itemize}

\vspace{2pt}
\textbf{9. Lack of Fact Verification/Logic Following.} This category includes cases where a reasoning LLM generates plausible but logically unsound reasoning by skipping necessary checks or failing to verify facts.

\textit{Example:} Claude produced incorrect results for the task \textit{HumanEval-59}. The number $13194$ is divisible by $2639$, but $2639$ is not prime ($2639 = 7 \times 13 \times 29$). The largest prime factor of $13195$ is $29$, but the loop stopped early at $2639$ instead of continuing to $29$ because it did not verify if $2639$ was prime.

% % Please add the following required packages to your document preamble:
% % \usepackage{multirow}
% \begin{table}[t!]
% \centering
% \caption{Distribution of the reasoning models' errors on the two datasets.\song{@M: you have four different LLMs, how this table were created?}\mamad{this table is aggregated results, the results for different llms are on the next table}} 
% \label{tab:failure_patterns}
% \setlength{\tabcolsep}{2pt}
% \scalebox{0.9}{
% \begin{tabular}{|c|lc|cc|cc|}
% \hline
% \multirow{2}{*}{Category} & \multicolumn{2}{c|}{regular inputs}& \multicolumn{2}{c|}{edge inputs} \\ \cline{2-5} 
% & \multicolumn{1}{c|}{HumanEval+} & LiveCodeBench &  \multicolumn{1}{c|}{HumanEval+} & LiveCodeBench \\ \hline
% Computation Errors & \multicolumn{1}{c|}{53}& 138 & \multicolumn{1}{c|}{133}& 343 \\ \hline
% Indexing Errors &  \multicolumn{1}{c|}{15}& 56 & \multicolumn{1}{c|}{25}& 33\\ \hline
% Control Flow Errors & \multicolumn{1}{c|}{4}& 31 & \multicolumn{1}{c|}{8}& 46  \\ \hline
% Skipping Statements & \multicolumn{1}{c|}{20}& 38 & \multicolumn{1}{c|}{30}& 27\\ \hline
% Misreporting Final Output & \multicolumn{1}{c|}{17}& 36 & \multicolumn{1}{c|}{49}& 17\\ \hline
% Input Misread & \multicolumn{1}{c|}{18}& 8 & \multicolumn{1}{c|}{32}& 39\\ \hline
% Native API Misevaluation & \multicolumn{1}{c|}{17}& 4 & \multicolumn{1}{c|}{28}& 6\\ \hline
% Hallucination & \multicolumn{1}{c|}{3}& 2 & \multicolumn{1}{c|}{20}& 9\\ \hline
% Lack of Fact Verification/Logic Following & \multicolumn{1}{c|}{5}& 31 & \multicolumn{1}{c|}{12}& 51\\ \hline
% \end{tabular}
% }
% \end{table}

\begin{table}[t!]
\centering
\caption{Distribution of reasoning failures for each LLM according to the identified failure patterns.}
\label{tab:failure_patterns}
\setlength{\tabcolsep}{2pt}
\scalebox{0.8}{
\begin{tabular}{|c|c|cc|cc|}
\hline
\multirow{2}{*}{Category} & \multirow{2}{*}{LLM} & \multicolumn{2}{c|}{Regular Inputs} & \multicolumn{2}{c|}{Edge Inputs} \\ \cline{3-6} 
 &  & HumanEval+ & LiveCodeBench & HumanEval+ & LiveCodeBench \\ \hline

\multirow{4}{*}{Computation Errors} 
 & Claude & 16 & 46 & 27 & 101 \\ 
 & DeepSeek & 20 & 62 & 28 & 114 \\ 
 & Gemini & 7 & 20 & 34 & 59 \\ 
 & GPT & 10 & 10 & 47 & 69 \\ \hline

\multirow{4}{*}{Control Flow Errors} 
 & Claude & 2 & 13 & 5 & 20 \\ 
 & DeepSeek & - & 9 & 1 & 11 \\ 
 & Gemini & 2 & 6 & 2 & 10 \\ 
 & GPT & - & 3 & - & 5 \\ \hline

\multirow{4}{*}{Hallucination} 
 & Claude & 1 & 1 & 6 & 3 \\ 
 & DeepSeek & 1 & 1 & 5 & - \\ 
 & Gemini & 1 & - & 6 & 4 \\ 
 & GPT & - & - & 3 & 2 \\ \hline

\multirow{4}{*}{Indexing Errors} 
 & Claude & 6 & 23 & 7 & 16 \\ 
 & DeepSeek & 4 & 12 & 8 & 5 \\ 
 & Gemini & 4 & 18 & 3 & 10 \\ 
 & GPT & 2 & 3 & 4 & 2 \\ \hline

\multirow{4}{*}{Input Misread} 
 & Claude & 8 & 1 & 9 & 9 \\ 
 & DeepSeek & 2 & 2 & 12 & 11 \\ 
 & Gemini & 8 & 5 & 8 & 6 \\ 
 & GPT & - & - & 3 & 13 \\ \hline

\multirow{4}{*}{Lack of Fact Verification/Logic Following} 
 & Claude & 1 & 7 & 1 & 10 \\ 
 & DeepSeek & 2 & 10 & 3 & 17 \\ 
 & Gemini & 1 & 8 & 5 & 15 \\ 
 & GPT & 1 & 6 & 3 & 9 \\ \hline

\multirow{4}{*}{Misreporting Final Output} 
 & Claude & 4 & 9 & 3 & 4 \\ 
 & DeepSeek & 2 & 12 & 13 & 4 \\ 
 & Gemini & 6 & 12 & 18 & 5 \\ 
 & GPT & 2 & 3 & 15 & 4 \\ \hline

\multirow{4}{*}{Native API Misevaluation} 
 & Claude & 4 & 3 & 10 & 1 \\ 
 & DeepSeek & - & 1 & 11 & 3 \\ 
 & Gemini & 8 & - & 2 & 2 \\ 
 & GPT & 5 & - & 5 & - \\ \hline

\multirow{4}{*}{Skipping Statements} 
 & Claude & 8 & 23 & 15 & 14 \\ 
 & DeepSeek & 5 & 9 & 9 & 9 \\ 
 & Gemini & 3 & 7 & 2 & 1 \\ 
 & GPT & 4 & - & 4 & - \\ \hline
\end{tabular}
}
\end{table}

Table~\ref{tab:failure_patterns} presents the distribution of reasoning failures for each LLM according to the identified failure patterns. 
The first observation is the dominance of computation errors across both datasets. These include arithmetic errors, type-conversion issues, and bitwise miscalculations. In HumanEval+, 53 failures under regular inputs and 133 under edge inputs were attributed to computation errors, while LiveCodeBench registered a substantially higher 138 and 343 failures in the same categories. This confirms that numerical reasoning and low-level operations are primary bottlenecks in LLM-based code inference.

%We can also observe that Indexing and control-flow errors increase with task complexity, with indexing errors rising from 15 to 56 and control-flow errors from 4 to 31 (regular inputs), reflecting LiveCodeBench’s emphasis on complex loops and branching. Skipping statements remains stable, while output misreporting spikes in HumanEval+ edge cases, indicating challenges in result presentation. Input misreads grow under edge conditions, whereas API misuse is more frequent in HumanEval+. Though less common, hallucinations and logic-following failures have a notable impact, with hallucinations rising under edge inputs and logic-following errors surging in LiveCodeBench, highlighting deeper reasoning challenges in complex tasks.
Indexing and control flow errors show more nuanced patterns. Indexing errors increase from 15 (HumanEval+ regular) to 56 (LiveCodeBench regular), reflecting the higher complexity of data manipulation tasks in LiveCodeBench. Interestingly, under edge inputs, indexing errors decrease from 25 (HumanEval+) to 33 (LiveCodeBench), suggesting that HumanEval+ may contain a higher concentration of corner cases stressing array boundaries. However, control flow errors scale sharply with the difficulty of the dataset, rising from 4 (HumanEval+ regular) to 31 (LiveCodeBench regular) and from 8 to 46 under edge inputs. These trends highlight the greater emphasis of LiveCodeBench on intricate loop and branch logic.

% \begin{table}[t!]
% \centering
% \caption{Error distribution across different LLMs.\song{make this as a latex figure}}
% \label{tab:error_distribution}
% \setlength{\tabcolsep}{2pt}
% \scalebox{0.95}{
% \begin{tabular}{|l|r|r|r|r|}
% \hline
% \textbf{Error Category} & \textbf{Gemini2.5 Flash} & \textbf{GPT-o4} & \textbf{Claude 4 Sonnet} & \textbf{DeepSeek R1} \\
% \hline
% Computation Error              & 120 & 136 & 190 & 224 \\
% Indexing Errors                & 35  & 11  & 52  & 28  \\
% Control Flow Errors            & 20  & 8   & 40  & 21  \\
% Skipping Statements            & 12  & 11  & 53  & 32  \\
% Misreporting Final Output      & 40  & 24  & 20  & 29  \\
% Input Misread                  & 26  & 16  & 24  & 27  \\
% Native API Misevaluation       & 11  & 10  & 18  & 15  \\
% Hallucination                  & 11  & 5   & 11  & 7   \\
% Lack of Fact Verification / Logic Following & 29 & 19 & 19 & 32 \\
% \hline
% \textbf{Total}                 & \textbf{304} & \textbf{240} & \textbf{427} & \textbf{415} \\
% \hline
% \end{tabular}
% }
% \end{table}

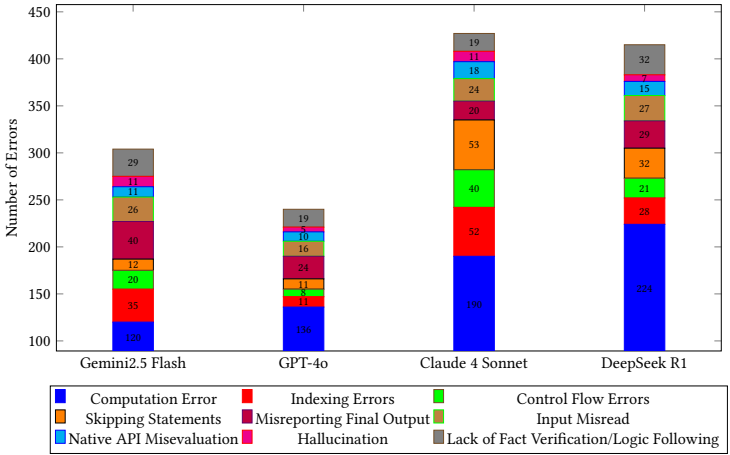
\begin{figure}[t!]
    \centering
    \resizebox{0.7\columnwidth}{!}{%
    \begin{tikzpicture}
        \begin{axis}[
            ybar stacked,
            bar width=25pt,
            width=16cm,
            height=9cm,
            ylabel={Number of Errors},
            symbolic x coords={Gemini2.5 Flash, GPT-4o, Claude 4 Sonnet, DeepSeek R1},
            xtick=data,
            enlarge x limits=0.15,
            nodes near coords,
            every node near coord/.append style={black, font=\scriptsize, yshift=0pt},
            legend style={at={(0.5,-0.1)}, anchor=north, legend columns=3},
        ]

        % Computation Error
        \addplot+[fill=blue, nodes near coords, nodes near coords style={black, font=\scriptsize, yshift=25pt}, point meta=explicit symbolic] coordinates {
            (Gemini2.5 Flash,120) [120]
            (GPT-4o,136) [136]
            (Claude 4 Sonnet,190) [190]
            (DeepSeek R1,224) [224]
        };

        % Indexing Errors
        \addplot+[fill=red, nodes near coords, point meta=explicit symbolic] coordinates {
            (Gemini2.5 Flash,35) [35]
            (GPT-4o,11) [11]
            (Claude 4 Sonnet,52) [52]
            (DeepSeek R1,28) [28]
        };

        % Control Flow Errors
        \addplot+[fill=green, nodes near coords, point meta=explicit symbolic] coordinates {
            (Gemini2.5 Flash,20) [20]
            (GPT-4o,8) [8]
            (Claude 4 Sonnet,40) [40]
            (DeepSeek R1,21) [21]
        };

        % Skipping Statements
        \addplot+[fill=orange, nodes near coords, point meta=explicit symbolic] coordinates {
            (Gemini2.5 Flash,12) [12]
            (GPT-4o,11) [11]
            (Claude 4 Sonnet,53) [53]
            (DeepSeek R1,32) [32]
        };

        % Misreporting Final Output
        \addplot+[fill=purple, nodes near coords, point meta=explicit symbolic] coordinates {
            (Gemini2.5 Flash,40) [40]
            (GPT-4o,24) [24]
            (Claude 4 Sonnet,20) [20]
            (DeepSeek R1,29) [29]
        };

        % Input Misread
        \addplot+[fill=brown, nodes near coords, point meta=explicit symbolic] coordinates {
            (Gemini2.5 Flash,26) [26]
            (GPT-4o,16) [16]
            (Claude 4 Sonnet,24) [24]
            (DeepSeek R1,27) [27]
        };

        % Native API Misevaluation
        \addplot+[fill=cyan, nodes near coords, point meta=explicit symbolic] coordinates {
            (Gemini2.5 Flash,11) [11]
            (GPT-4o,10) [10]
            (Claude 4 Sonnet,18) [18]
            (DeepSeek R1,15) [15]
        };

        % Hallucination
        \addplot+[fill=magenta, nodes near coords, point meta=explicit symbolic] coordinates {
            (Gemini2.5 Flash,11) [11]
            (GPT-4o,5) [5]
            (Claude 4 Sonnet,11) [11]
            (DeepSeek R1,7) [7]
        };
        % Lack of Fact Verification / Logic Following
        \addplot+[fill=gray, nodes near coords, point meta=explicit symbolic] coordinates {
            (Gemini2.5 Flash,29) [29]
            (GPT-4o,19) [19]
            (Claude 4 Sonnet,19) [19]
            (DeepSeek R1,32) [32]
        };
        \legend{
            Computation Error,
            Indexing Errors,
            Control Flow Errors,
            Skipping Statements,
            Misreporting Final Output,
            Input Misread,
            Native API Misevaluation,
            Hallucination,
            Lack of Fact Verification/Logic Following
        }
        \end{axis}
    \end{tikzpicture}
    }
    \caption{The distribution of error categories per LLM. Each bar represents an LLM, and each segment shows the contribution of a specific error category.}
    \label{fig:llm_error_distribution_stacked}
\end{figure}

Skipping statements and misreporting final outputs further reveal weaknesses in execution fidelity. Skipping-related errors appear in both datasets but remain more stable in all conditions (20–38 cases in regular inputs, 27–30 in edge inputs). However, if the final output is misreported, there are spikes in HumanEval + edge cases (49), compared to only 17 in LiveCodeBench edge cases. This indicates that in smaller datasets like HumanEval+, models may succeed in reasoning steps but fail at correctly presenting the final result, whereas LiveCodeBench errors are more deeply rooted in the reasoning process itself.

Input misreads and API misevaluation show contrasting profiles. Input misreading errors grow sharply under edge inputs (32 in HumanEval+, 39 in LiveCodeBench), reflecting difficulties in parsing irregular or length-sensitive inputs. By contrast, Native API misevaluation is relatively rare in LiveCodeBench (4–6 errors), compared to higher counts in HumanEval+ (17–28). This suggests that HumanEval+ contains more problems where reliance on library defaults or Python-specific semantics plays a decisive role.

Hallucination and lack of fact verification/logic are relatively less frequent but notable for their qualitative impact. Hallucination errors, such as fabricating runtime exceptions or numerical values, remain limited in frequency but increase under edge conditions (20 for HumanEval+, 9 for LiveCodeBench). More concerning is the category of logic-following failures, which grows dramatically in LiveCodeBench: from 5 (HumanEval+ regular) to 31, and from 12 (HumanEval+ edge) to 51. This highlights that beyond surface-level reasoning, LiveCodeBench systematically stresses deeper logical consistency, leading to a greater concentration of higher-level reasoning failures.

We also show the distribution of failure occurrences across the categories in Figure~\ref{fig:llm_error_distribution_stacked},  for the four reasoning LLMs we evaluated. Overall, Claude 4 exhibits the highest number of failures with 427, followed by DeepSeek R1 with 415. Gemini and GPT-o4 produce roughly 50\% fewer failures. 

\mybox{\textbf{Answer to RQ3:} 
The inference failures of reasoning LLMs can be categorized into nine distinct types:  \textit{Computation Errors}, \textit{Indexing Errors}, \textit{Control Flow Errors}, \textit{Skip Statements}, \textit{Misreporting Final Output}, \textit{Input Misread}, \textit{Misevaluation of Native API}, \textit{Hallucination}, and \textit{Lack of Verification/Logic Following}, which help capture the diverse ways in which models produce incorrect inference outputs. \textit{Computation Errors} are the most dominant failure patterns, followed by \textit{Control Flow Errors} and \textit{Indexing Errors}.}

\section{Discussion}
\label{sec:discussion}
This section addresses the open questions of our study. In Section~\ref{sec:5.1}, we examine whether code complexity affects LLM reasoning performance. In Section~\ref{sec:5.2}, we investigate whether a too-argumented approach can improve LLM reasoning performance in inferring code execution results. 

\subsection{Reasoning Errors vs. Code Complexity}
\label{sec:5.1}
In this section, we examine whether code complexity affects LLM reasoning performance. Specifically, we use the Mann-Whitney U test to determine if there are significant differences in metrics such as CC, HM, MI, and LOC between code samples that the LLM successfully solved and those it failed to solve.

% \begin{table}[t!]
% \centering
% \caption{Overview of the Distribution of Code Complexity Across Error Categories. \textbf{CC} denotes the Cyclomatic Complexity metric, \textbf{HM} is the Halstead Metrics (difficulty), and \textbf{MI} is the Maintainability Index.}
% \label{tab:error_code_complexity}

% \resizebox{\textwidth}{!}{%
% \begin{tabular}{|l|c|c|c|c|c|}
% \hline
% \textbf{Dataset}  & \textbf{\# Unique Tasks} & \textbf{Avg. \#LOC} & \textbf{Avg. CC}  & \textbf{Avg. HM} & \textbf{Avg. MI} \\ \hline
% Computation Errors & 249 & 11.90 & 2.74 & 2.32 & 72.14 \\ \hline
% Indexing Errors & 76 & 12.79 & 2.71 & 2.17 & 69.36 \\ \hline
% Control Flow Errors & 69 & 17.10 & 3.95 & 3.44 & 66.14 \\ \hline
% Skipping Statements & 66 & 13.77 & 3.18 & 2.63 & 69.74 \\ \hline
% Misreporting Final Output & 62 & 12.90 & 2.71 & 1.89 & 69.89 \\ \hline
% Input Misread & 62 & 12.39 & 2.89 & 1.94 & 73.20 \\ \hline
% Native API Misevaluation & 16 & 10.17 & 2.83 & 1.24 & 76.57 \\ \hline
% Hallucination & 23 & 15.38 & 4.00 & 3.09 & 67.35 \\ \hline
% Lack of Fact Verification/Logic Following & 62 & 16.33 & 3.92 & 3.40 & 67.39 \\ \hline
% \end{tabular}
% }
% \end{table}

\begin{table}[t!]
\centering
\caption{Mann–Whitney U test comparing code complexity metrics between tasks where LLM reasoning failed and those where it succeeded. 
The last two columns show the average metric values for tasks with reasoning failures and successful reasoning, respectively.}
\label{tab:complexity_mannwhitney}
\scalebox{0.88}{
\begin{tabular}{|l|c|c|c|c|}
\hline
\textbf{Complexity Metric} & \textbf{U-statistic} & \textbf{p-value} & \textbf{Avg. (Failed Tasks)} & \textbf{Avg. (Successful Tasks)} \\ \hline
Lines of Code (LOC)      & 22\,274.0 & 0.4025 & 11.69 & 10.55 \\ \hline
Halstead Difficulty (HD)  & 16\,561.5 & 0.001959 & 2.21  & 2.84  \\ \hline
Cyclomatic Complexity (CC)& 12\,504.5 & 2.40$\times$10$^{-9}$ & 2.78  & 4.10  \\ \hline
Maintainability Index (MI)& 26\,557.5 & 0.00015575 & 72.96 & 68.15 \\ \hline
\end{tabular}
}
\end{table}

Based on the results of the Mann–Whitney U test results shown in Table~\ref{tab:complexity_mannwhitney}, we draw several insights into the relationship between code complexity and LLM reasoning success. 
First, cyclomatic Complexity (CC) and the Maintainability Index (MI) exhibit strong and statistically significant differences between solved and failed tasks, suggesting that reasoning LLMs more often succeed on structurally complex, less maintainable code. Second, Lines of Code (LOC) does not show a statistically significant difference, indicating that program size alone is not a reliable predictor of reasoning accuracy. Finally, Halstead Difficulty (HD) reveals an intriguing pattern: tasks that trigger reasoning failures tend to involve less complex code than tasks where models succeed. This implies that reasoning LLMs are not merely more error-prone on cognitively demanding programs, but may also struggle unexpectedly with seemingly straightforward ones.

\subsection{Can LLM Reasoning be Enhanced through Tool-augmented Reasoning?}
\label{sec:5.2}

In this study, we have demonstrated that reasoning LLMs can produce incorrect inference results for code execution tasks, and mitigating these failures requires different types of effort depending on the error category. For example, \textit{Computation Errors} and \textit{Indexing Errors} demand more precise calculations, which external computational tools can support. Inference errors related to \textit{Control Flow Errors} and \textit{Skipped Statements} require more rigorous program analysis, leveraging AST/CFG-based reasoning to track structural dependencies. \textit{Native API Misevaluation} calls for actual invocation and execution of the APIs to obtain reliable outcomes. Failures like \textit{Misreporting of Final Output} and \textit{Input Misreads} can be alleviated through strict pre- and post-processing, ensuring consistent formatting and robust input parsing. Finally, \textit{Hallucinations} and \textit{Lack of Verification/Logic Following} underscore the need for more robust verification mechanisms, thorough cross-checking, and possibly external logic validators to ensure coherence and correctness.

In this section, we focus on the most prevalent failure category, i.e., \textit{Computation Errors}, and propose a tool-augmented reasoning approach to examine whether external tools can improve LLM reasoning ability. We first constructed an experimental dataset by randomly sampling, with 95\% confidence, from the original task set containing such errors. The sampled cases were drawn from both HumanEval+ and LiveCodeBench, covering failures under both regular and edge inputs. For each instance, we used the trace-level failures identified in Section~\ref{sec:3.4} as a starting point. We then manually created a simple Python mathematical clause that correctly implemented the intended computation, which the LLM had previously miscomputed, and executed it to obtain the accurate result. Finally, we fed the correct segment of the reasoning chain together with the verified result back to the model, prompting it to update its reasoning and regenerate the final answer. This process not only corrects specific computational mistakes but also illustrates how tool augmentation can systematically guide LLMs toward more reliable and accurate reasoning.

\begin{table}[t!]
\centering
\caption{Impact of tool-augmented reasoning on \textit{Computation Errors}}
\label{tab:tool_aug_results}
\scalebox{0.9}{
\begin{tabular}{|l|c|c|}
\hline
\textbf{LLM Model} & \textbf{\# errors w/o tool-augmentation} & \textbf{\# errors with tool-augmentation} \\ \hline
Gemini 2.5     & 82  & 36  \\ \hline
GPT-4o mini         & 71  & 32  \\ \hline
Claude 4 Sonnet      & 131 & 56  \\ \hline
DeepSeek R1    & 155 & 57 \\ \hline \hline
\textbf{Total} &439&181 \\ \hline
\end{tabular}
}
\end{table}

Table~\ref{tab:tool_aug_results} reports the number of errors produced by each LLM in the \textit{Computation Errors} category, both before and after applying our tool-augmented reasoning approach for code result inference. As shown, the proposed pipeline substantially reduces computation errors across all models. Each LLM benefits consistently, achieving reductions of approximately 50–60\% in observed errors. Overall, our approach resolves 58\% of the original inference failures attributed to \textit{Computation Errors}, demonstrating the effectiveness of tool augmentation in strengthening LLM reasoning.
%\song{give some reasons why some of the errors cannot be removed with a tool-argumented approach: revise the following draft}

Nevertheless, approximately 40\% of errors in the \textit{Computation Errors} category remain uncorrected even with our tool-augmented approach. These residual failures arise from several causes. In some cases, the model fails to properly integrate the externally computed results into its reasoning chain. In other cases, the errors are compounded by higher-level reasoning flaws, such as incorrect loop or branch handling, or the propagation of earlier mistakes, that cannot be resolved by simply supplying the correct numerical value. We also observe instances where new errors are introduced during reasoning, even when the model is provided with the correct calculation, suggesting instability in how LLMs revise their inference process. 
Additionally, some failures stem from limitations in the model’s ability to maintain context or to align tool outputs with the intended program semantics. Together, these factors highlight that while tool augmentation effectively mitigates straightforward computation mistakes, more complex reasoning errors require deeper integration of program analysis and verification mechanisms. 

These results also indicate that tool-augmented reasoning can be extended beyond this experiment. A promising next step is to design an agent that automatically invokes external tools during code inference or generation, enabling scalable and more reliable tool-augmented generation. 

\section{Threats To Validity}
\label{sec:threats}
\textbf{External validity} concerns whether our findings extend beyond the particular setting of this study. Our analysis was carried out on reasoning LLMs using Python code inference tasks. Although Python is a natural choice given its dominance in research and practice, the conclusions may not apply directly to other programming languages with different characteristics, such as strict typing, memory management, or concurrency models. To gain a broader perspective, future studies should replicate our methodology in other ecosystems (e.g., Java, C++, JavaScript). Since our evaluation procedure is not tied to a specific language, it can be readily adapted, which would strengthen the generalizability of the results.  

\textbf{Internal validity} threats stem mainly from the human-driven parts of the study, particularly the manual error annotation and taxonomy development. Classifying reasoning errors can be subjective, and annotators may occasionally disagree on how to interpret ambiguous traces. To mitigate this, we defined clear guidelines for each error category, and two authors independently reviewed the annotations. Discrepancies were resolved through discussions until agreement was reached. Although this process improved reliability, some subjectivity is inevitable in qualitative analysis.  

\textbf{Construct validity} relates to whether our evaluation truly measures the reasoning failures we aim to capture. The results may be influenced by choices in prompt wording, dataset format, or the execution environment itself. We reduced these risks by experimenting with multiple prompt formulations and selecting the one that produced consistent outputs across models.
\section{Conclusion}
\label{sec:conclusion}

In this study, we conduct the first empirical investigation of reasoning runtime behavior in LLMs, focusing on code execution inference. We curated a benchmark of 427 code snippets from HumanEval+ and LiveCodeBench, evaluating four state-of-the-art reasoning LLMs (including DeepSeek-R1, OpenAI o4-mini, Gemini 2.5 Flash, and Claude 4 Sonnet) across three input types, i.e., regular, edge, and invalid, with each snippet paired with 12 ground-truth input-output values. Our findings show that these models can achieve high overall accuracies (85–98\%). We further analyze the reasoning errors and propose a comprehensive taxonomy of nine inference error types, offering a structured understanding of LLM limitations. To mitigate these errors, we explore a tool-augmented reasoning approach focused on \textit{Computation Errors}, demonstrating that it can correct 58\% of such failures. These results highlight the promise of integrating external tools with LLMs to enhance reasoning reliability, improve execution inference, and provide a pathway for systematically addressing diverse failure modes.

\section{Data Availability}
We release the data and source code of our experiments to enable other researchers to replicate and extend our study
(\url{https://anonymous.4open.science/r/Demystifying_the_Reasoning_Errors_of_LLM-8597/README.md}).
%\song{create an 4open.science repo to share 1) your data, 2) script for reach RQ, make sure we can rerun your script to replicate the results, 3. readme about how to run the scripts;} \song{check the example:  \url{https://anonymous.4open.science/r/api_guided_testgen-FB88/README.md}}
%\todo{Need to add a statement on shared artifact; Does not take up the page limit; }

\bibliographystyle{ACM-Reference-Format}

\bibliography{paper}

% \appendix

\end{document}